\documentclass[conference,a4paper,final]{IEEEtran}
\IEEEoverridecommandlockouts
\usepackage{cite}
\usepackage{amsmath,amssymb,amsfonts}
\usepackage{graphicx}
\usepackage{subfigure}
\usepackage{url}
\usepackage{textcomp}
\usepackage{xcolor}
\usepackage{soul,color}
\usepackage{bm}
\usepackage{booktabs}
\usepackage{array,multirow,graphicx}
\usepackage{algorithm}
\usepackage{algpseudocode}
\usepackage[top=2.0cm,bottom=4.4cm,left=1.4cm, right=1.4cm]{geometry}

\usepackage{mdwlist}
\usepackage{neuralnetwork}
\usepackage{balance}

\begin{document}
\title{Detecting 5G Signal Jammers Using Spectrograms with Supervised and Unsupervised Learning\thanks{This work was partially supported by the German Federal Office for Information Security within the project ADWISOR5G under grant ID 01MO23030B. This work has been partially funded by the European Commission through the Horizon Europe/JU SNS project ROBUST-6G (Grant Agreement no. 101139068).}}

\author{
\IEEEauthorblockN{Matteo Varotto$^\star$, Stefan Valentin$^\star$, and Stefano Tomasin$^{\star\star}$}\smallskip
\IEEEauthorblockA{\url{matteo.varotto@h-da.de}, \url{stefan.valentin@h-da.de}, \url{tomasin@dei.unipd.it}}\medskip
\IEEEauthorblockA{
$^{\star}$Dep. of Computer Science, Darmstadt University of Applied Sciences, Germany\\
$^{\star\star}$Dep. of Information Engineering, Dep. of Mathematics, University of Padova, Italy
}
}

\maketitle

\begin{abstract}
Cellular networks are potential targets of jamming attacks to disrupt wireless communications. Since the fifth generation (5G) of cellular networks enables mission-critical applications, such as autonomous driving or smart manufacturing, the resulting malfunctions can cause serious damage. This paper proposes to detect broadband jammers by an online classification of spectrograms. These spectrograms are computed from a stream of in-phase and quadrature (IQ) samples of 5G radio signals. We obtain these signals experimentally and describe how to design a suitable dataset for training. Based on this data, we compare two classification methods: a supervised learning model built on a basic convolutional neural network (CNN) and an unsupervised learning model based on a convolutional autoencoder (CAE). After comparing the structure of these models, their performance is assessed in terms of accuracy and computational complexity.
\end{abstract}


\section{Introduction}
The advent of fifth-generation (5G) technology promises very high data rates, low latency, and the support of mission-critical applications. However, 5G networks are vulnerable to jamming attacks which may cause a denial of service (DoS) of critical applications, with potentially serious consequences on persons and things \cite{survey}.

One approach to cope with the threat of jamming is the use of wireless intrusion prevention systems (WIPSs) that monitor communication by analyzing features such as packet error rate (PER), bit error rate (BER), and signal-to-interference-plus-noise ratio (SINR) \cite{zhang2010overview}. Using such features at a relatively high abstraction level (i) may be misleading since their high variation is typical in wireless channels and can, thus, only hardly be attributed to a single cause and (ii) has been shown to fail at detecting jammers that target essential 5G signaling channels, such as the signal synchronization block (SSB) \cite{jamming_strategies}. At a lower abstraction level, we find approaches that manipulate the 5G radio signals, e.g., by nulling some subcarriers and comparing the received power on such subcarriers with a threshold \cite{jamm_det,jamm_det_2,9569600}. This not only lowers the data rate of the system but also requires changes in current cellular network standards and systems. It is also inefficient since a simple threshold can be easily evaded by an intermittent jammer \cite{9569600}.

From a methodological perspective, some early machine learning (ML) and deep learning (DL) models have shown promising results through the direct analysis of received radio signals \cite{5g_ml,jamming_ml}. Effective features were the number of transmissions or the clear channel assessments \cite{hachimi} or aggregate measurements on the link layer \cite{jere2023machine}.

In this paper, we propose a WIPS that obtains features directly from the radio signal at the physical baseband. Based on received in-phase and quadrature (IQ) samples, a stream of spectrograms is computed, which is then used by a machine learning model to detect jammed signals. This process can be performed on a separate system (called watchdog) and requires neither changes to the 5G architecture nor to its signals. The watchdog can be functionally simple as measuring received power requires only static parameterization, without further processing the radio signals, e.g., for equalization or decoding. A spectrogram, or more precisely, a power spectral density (PSD), can be still obtained from power measurements even when the received signal power is too low for communication. This allows to detect jamming attacks even at very low SINR -- an important benefit compared to the mentioned approaches based on specific OFDM signals \cite{jamm_det,jamm_det_2,9569600} and to the approaches using link-layer measurements \cite{zhang2010overview,hachimi,jere2023machine}. 

Using spectrograms as samples, we will compare a (i) convolutional neural network (CNN) trained with \textit{supervised} learning to a (ii) convolutional autoencoder (CAE) built with \textit{unsupervised} learning in terms of accuracy and computational complexity. Unsupervised learning is only using data from the non-jammed system, which avoids overfitting the model to a specific jammer. Supervised learning, however, requires data from jammed cases. We produce this data from lab experiments based on software-defined radios (SDRs), where a broadband jammer attacks an indoor 5G network.

For this scenario, we will experimentally demonstrate high detection rates and short classification times. After detailing the scenario in Sec. \ref{sec:scenario}, we describe experimental setup and data collection in Sec. \ref{sec:data}. Designing and training the machine learning models are described in Sec. \ref{sec:models} and performance results are presented in Sec. \ref{sec:results}. Sec. \ref{sec:concl} concludes the paper.

\section{Security Scenario}
\label{sec:scenario}
\begin{figure}
\includegraphics[width=1.0\hsize]{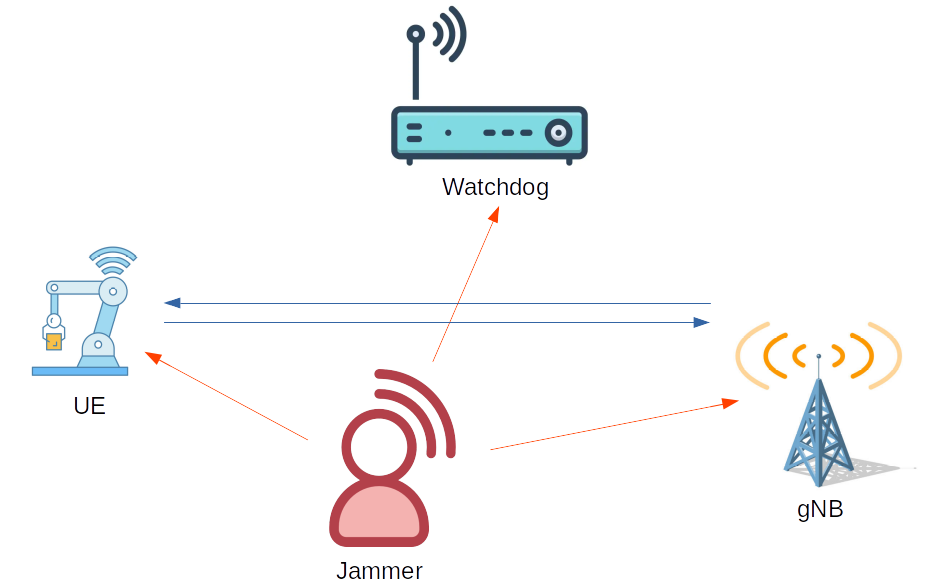}
\caption{Considered security scenario: blue arrows indicate legitimate cellular communications and red arrows indicate the jamming signals.}
\label{scenario}
\end{figure}

We consider the scenario in Fig. \ref{scenario}. We assume that the area of interest is served by at least one legitimate cellular network. This setting includes the case of private 5G networks for industrial applications. In such a scenario, cellular communications are used to support industrial activity, e.g., to connect robots and production devices, or to coordinate devices and operators as in railway networks \cite{5g_iot}.
\newline
The attacker aims to disrupt the network's operation until Denial of Service (Dos) is achieved. We focus on the jamming attack, where the attacker transmits a signal that interferes with regular cellular communications and prevents the correct decoding of received data. To this end, the attacker may simply transmit a powerful noise-like signal to reduce the SINR.

A first defense against jamming is its detection, which enables countermeasures such as the localization and removal of the jammer. Focusing on detection only, we consider the presence of a dedicated device, called \emph{ watchdog}, which detects the presence of jammed radio signals. The watchdog is not associated with the monitored radio access network (RAN), i.e., it does not exchange information with the base station (gNB).

Without excluding control through higher-layer networks (e.g., core network), this separation simplifies the watchdog deployment in many contexts. For instance, users or operators that are not under the direct control of their 5G network (or have doubts thereof) may deploy a watchdog without further notice. Such a case is not unlikely in the fragmented business space of mobile networks, where network operators have outsourced most of their RAN operation and deployment. It is worth noting that a watchdog only receives wireless signals, thus remaining undetectable to the attacker at the radio level.

\section{Dataset Creation}
\label{sec:data}

\subsection{Basic assumptions}
We assume that the watchdog knows basic radio parameters. This assumption is feasible since the required numerology (i.e., carrier frequency, bandwidth, and pilot structure) is constant and known to the RAN operator. With this parameterization, the watchdog records data from the physical baseband channel, outputting a stream of IQ samples. From that stream, the watchdog obtains the \textit{spectrogram}, by first taking the fast Fourier transform (FFT) over a window of IQ samples and then obtaining the PSD array collecting the modulus square of each frequency sample. Then, $n$ PSD arrays are stacked into a final matrix, called spectrogram. This procedure is specified in Algorithm \ref{algo1}.
\begin{algorithm}[t]
\caption{Pseudocode for creating the spectrogram matrix} \label{algo1}
\small
\begin{algorithmic}
\Require sampling\textunderscore rate, IQ\textunderscore STREAM
\State mat$\gets$ zeros$(100,1024)$\Comment empty matrix
\State x $\gets$ READ\textunderscore IQ\textunderscore STREAM \Comment{Array for recorded IQ stream}
\State n $\gets 1024$ \Comment{Time window}
\State lower\textunderscore index $\gets 0$
\State upper\textunderscore index $\gets$ n 
\For{i$\gets 0$ to $100$}
\State y$\gets$ x[lower:upper] \Comment{Portion of the dataset of $1024$ samples.}
\State PSD$\gets$ FFT(y)$^{2}$ / n*sampling\textunderscore rate \Comment{Apply FFT to y}
\State PSD\textunderscore shifted $\gets$ FFT\textunderscore SHIFT(PSD) \Comment{Center PSD at 0 Hz}
\State mat[i:]$\gets$ PSD\textunderscore shifted \Comment{Insert PSD as i\textsubscript{th} row of the matrix}
\State lower\textunderscore index $\gets$ lower\textunderscore index$+$n
\State upper\textunderscore index $\gets$ upper\textunderscore index$+$n \Comment{Iterating over dataset}
\EndFor
\end{algorithmic}
\end{algorithm}

The collected IQ samples are assumed to contain three cases:
\begin{enumerate*}
    \item \textbf{Empty channel, no jammer:} gNB actively transmitting beacons but UE not transmitting,
    \item \textbf{Active channel, no jammer:} UE and\slash{}or gNB are transmitting data in time-division duplexing (TDD) mode,
    \item \textbf{Active Jammer:} UE and gNB occasionally send signals (e.g., beacons, connection requests) but no communication is possible.
\end{enumerate*}
The first two cases are classified as legitimate, while the third one is considered as anomalous. We assume that the jammer always sucessfully disrupts the communication between UE and gNB. This assumptions holds for jammers in close distance to the gNB, as verified experimentally in our laboratory scenario.

\subsection{Experimental Setup}
\begin{figure}
\includegraphics[width=1\hsize]{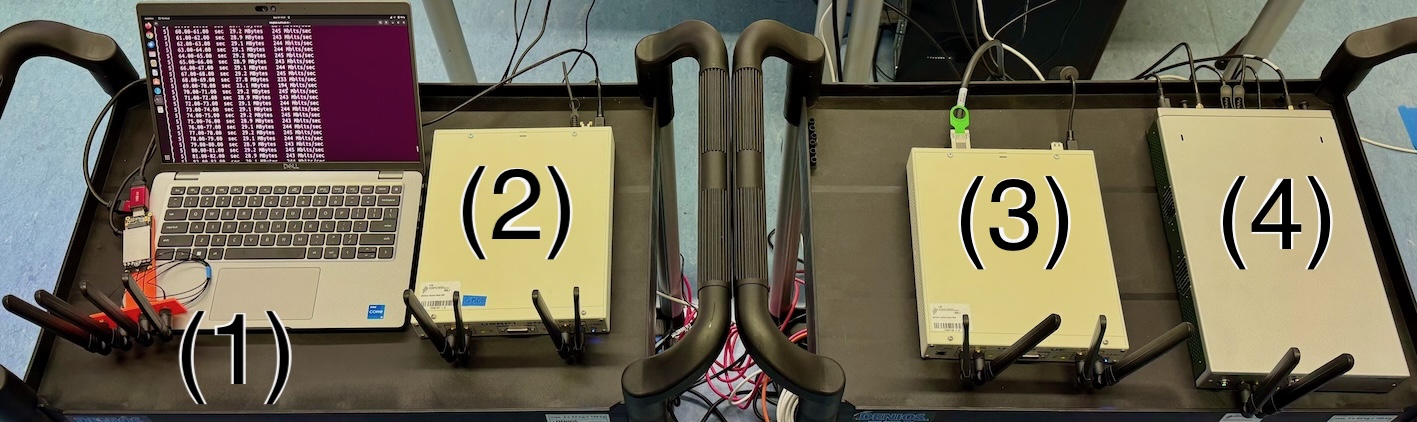}
\caption{Experimental setup: (1) 5G UE and the RF frontends for the (2) jammer, (3) watchdog, and (4) gNB. The corresponding PCs for (2--4) are not shown. The shown distances between the devices are for illustration purposes only. During experiments, the distance between adjacent devices was $1$\,m.}
\label{setup}
\end{figure}

Fig. \ref{setup} shows the experimental setup. We are running a private 5G network in the frequency band n78 with center frequency $f_c = 3750$~MHz. The system operates at 100\,MHz bandwidth in TDD mode. The gNB implements the 5G new radio (NR) air interface using srsRAN 23.10\cite{srsran} and the universal software radio peripheral (USRP) n300 radio frequency (RF) frontend \cite{usrp}. The user equipment (UE) is a Quectel RM520N-GL modem \cite{queltec}, which is connected via USB 3.0 to a laptop computer. The core network functionality is provided by Open5GS 2.6.6 \cite{open5gs}, running on the same
generic computer as srsRAN. This setting provides a 5G standalone network and complies
with Release 17.4.0 of the 3rd generation partnership project (3GPP) standard series 38 \cite{3gpp}. 

The jammer and the watchdog run on separate computers, each using one USRP X310 \cite{usrpx} RF frontend. The jammer permanently transmits uniform or Gaussian noise signals of $100$\,MHz bandwidth, thus covering the complete frequency band. The watchdog permanently records IQ samples over $120$\,MHz to include the potential emission into the neighboring bands.

Dataset creation, training and the measuring classification time was performed on a single workstation with an Intel Xeon w7-2495X CPU and an NVIDIA RTX A6000 GPU.

\subsection{Data processing}
An PSD array is obtained by applying an FFT to a window of $1024$ samples, leading to a frequency resolution of $117$\,kHz. We used an FFT rather than Welch's method \cite{Welsch}, thus, sacrificing precision for computational speed. A spectrogram is then composed as a $100\times 1024$ matrix by stacking $100$ PSDs. This corresponds to a time window of $0.8$~ms.

The resulting matrices presented two problems when fed into a deep learning (DL) model. First, the power of received radio signals is very low and is, thus, usually expressed in the logarithmic domain (decibel). Similarly, we apply the monotonic function $f(x)=-\log x$ to the value $x$ of PSD at each frequency, which avoids the vanishing gradient problem \cite{vanishing_gradient}. Second, due to approximation errors in the empty channel case, some values of the PSDs turned out to be $0$, which causes computation errors. We replaced such values by the small constant $\epsilon=10^{-21}$, leading to the applied transformation $f(x)=-\log (x+\epsilon)$.

\section{Modeling}
\label{sec:models}
We now describe the two models used to detect jammers, based on unsupervised and supervised learning.

\subsection{Unsupervised Learning}
With this approach, we adopt a CAE to detect jamming since the autoencoder (AE) is widely used for the one-class classification required for anomaly detection \cite{anomaly_det}. As we are dealing with two-dimensional (2D)
data structures with spatial correlation, convolution and max-pooling operations
are useful.

An AE is a DL structure that, given an input $\mathbf{X}$, compresses it to a latent space with reduced dimensionality (Encoder) and then reconstructs the original input by outputting $\mathbf{Y}$ (Decoder). In this case, $\mathbf{X}$ and $\mathbf{Y}$ are the $100\times 1024$ matrices obtained by PSD stacking. The goal of our model is then to minimize the mean squared error (MSE) as the loss function between the input and the output, i.e., 
\begin{equation}
\bar{\Gamma} = {\mathbb E}[\Gamma], \quad \Gamma = ||\bm{X} - \bm{Y}||^2.
\end{equation}
When the model is trained with \emph{no-jamming} spectrograms, it will minimize reconstruction error only in this case. For the \emph{jammed} cases, however, the reconstruction error can be assumed to be relatively high. In fact, when the CAE input corresponds to a signal including jamming, we expect that it will have a structure that cannot be properly described by the latent representation. 
\begin{table}
  \centering
  \renewcommand{\arraystretch}{1}
  \caption{Structure of the employed CAE}
  \begin{tabular}{c|llr}
    \toprule
    & Layer & Output size & No. of parameters\\
     \midrule
\parbox[t]{2mm}{\multirow{5}{*}{\rotatebox[origin=c]{90}{Encoder}}} &     Input              & $100 \times 1024 \times 1$ & $0$\\
     & Convolutional 1    & $49 \times 511 \times 32$ & $320$\\
     &Max Pooling   & $24 \times 255 \times 32$ & $0$\\
     &Convolutional 2           & $22\times 253\times 64$ & $18496$\\
     &Flatten              & $88704$ & $0$\\
     &Dense              & $8$ & $709640$\\
     \midrule
\parbox[t]{2mm}{\multirow{5}{*}{\rotatebox[origin=c]{90}{Decoder}}} &        Input                  & $8$ & $0$\\
&     Dense                  & $88704$ & $798336$\\
&     Reshape                & $11\times 126\times 64$ & $0$\\
&     Convolutional 1$^T$    & $23\times 253\times 128$ & $73856$\\
&     Convolutional 2$^T$    & $47\times 507\times 64$ & $73792$\\
&     Zero Padding             & $50\times 512\times 64$ & $0$\\
&     Convolutional 2$^T$    & $100\times 1024\times 1$ & $577$\\
    \bottomrule
  \end{tabular}
  \label{cae_par}
\end{table}
\newline
Letting $\mathcal H_0$ be the hypothesis class of no jamming and $\mathcal H_1$ the hypothesis class of jamming, the detection of jamming is performed by the following {\em test function} on the input image $\bm{X}$ to obtain decision $\hat{\mathcal H}$:
\begin{equation}
\hat{\mathcal H} = \begin{cases}
\mathcal H_0; & \Gamma < \tau, \\
\mathcal H_1; & \Gamma \geq \tau,\\
\end{cases}
\end{equation}
where $\tau$ is a chosen threshold. With this approach, type I and type II errors can be directly obtained. We, thus, measure accuracy as probability of false alarm (FA) and misdetection (MD) defined as
\begin{equation}
P_{\rm FA, C} = {\mathbb P}[\hat{\mathcal H} = \mathcal H_1 |\mathcal H = \mathcal H_0],
\end{equation}
\begin{equation}
P_{\rm MD, C} = {\mathbb P}[\hat{\mathcal H} = \mathcal H_0 |\mathcal H = \mathcal H_1].
\end{equation}

\subsection{Supervised Learning}
\begin{table}
  \centering
  \renewcommand{\arraystretch}{1}
  \caption{Structure of the employed CNN.}
  \begin{tabular}{c|llr}
    \toprule
    & Layer & Output size & No. of parameters\\
     \midrule
\parbox[t]{2mm}{\multirow{5}{*}{\rotatebox[origin=c]{90}{}}} &     Input              & $100 \times 1024 \times 1$ & $0$\\
     & Convolutional 1    & $49 \times 511 \times 32$ & $320$\\
     &Max Pooling   & $24 \times 255 \times 32$ & $0$\\
     &Convolutional 2           & $22\times 253\times 64$ & $18496$\\
     &Max Pooling & $11\times 126\times 64$ & $0$\\
     &Convolutional 3             & $9\times 124\times 128$ & $73856$\\
     &Flatten        & $31744$ & $0$\\
     &Dense        & $16$ & $507920$\\
     &Dense        & $8$ & $136$\\
     &Dense        & $1$ & $9$\\
     \bottomrule
  \end{tabular}
  \label{cnn_par}
\end{table}
While the AE only has to be trained with not-jammed cases in order to detect anomalies, with supervised learning this approach changes. First, the model is trained with jammed and not-jammed samples. To each sample $i$, a label $y_{i}$ is assigned, taking values $0$ for trusted cases and $1$ for jammed cases. Second, the samples are not reconstructed from a latent space but the output of the network is a single neuron. Third, the prediction is not computed on the same input sample but on the label associated with it.

The designed model is a CNN (recall that the input is a $2$D data structure) with the objective to return a $0$ at the final layer whenever the input sample is taken from the trusted cases and a $1$ whenever the sample is a jammed case. The chosen loss function is the binary cross-entropy, defined as
\begin{equation}
L = -\frac{1}{N}\sum_{i=1}^{N}y_{i}\cdot \log(\tilde{y}_{i})+ (1-y_{i})\cdot \log(1-\tilde{y}_{i}),
\end{equation}
where $\tilde{y}_{i}$ is the prediction of the $i$th sample. This function can approach infinity even if the prediction error cannot be above $1$, thus, allowing so the model to update its weights. 

For consistency, we also assess the performance of the supervised approach in terms of MD and FA as well as in computational complexity. Being a binary classification problem, MD and FA rates can be directly computed from the output of the last neuron of the CNN for a varying threshold.

\section{Performance results}
\label{sec:results}
\subsection{Unsupervised Learning}
\label{ssec:unsup_res}
The training set is composed of $6000$ matrices taken from trusted situations, with cases divided equally between an empty channel and an ongoing transmission. The validation set is used to monitor and to stop the training whenever the loss increases for $6$ epochs. This set is composed of $800$ matrices with the same equal distribution as the training set. The test set is composed of $800$ samples, divided equally between jammed and not-jammed cases.

\begin{figure}
    \centering
    \includegraphics[width=1.0\hsize]{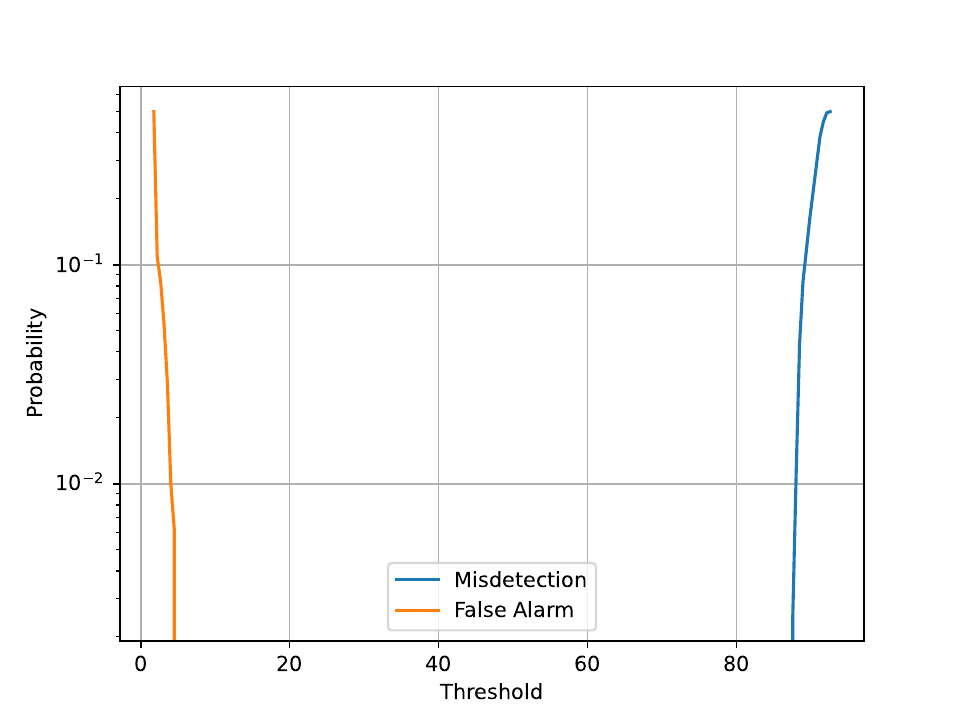}
    \caption{FA and MD probabilities as a function of the threshold $\tau$ for the uniform noise generator with the unsupervised learning approach.}
    \label{fig:unsup_error_uniform}
\end{figure}

\begin{figure}
    \centering
    \includegraphics[width=1.0\hsize]{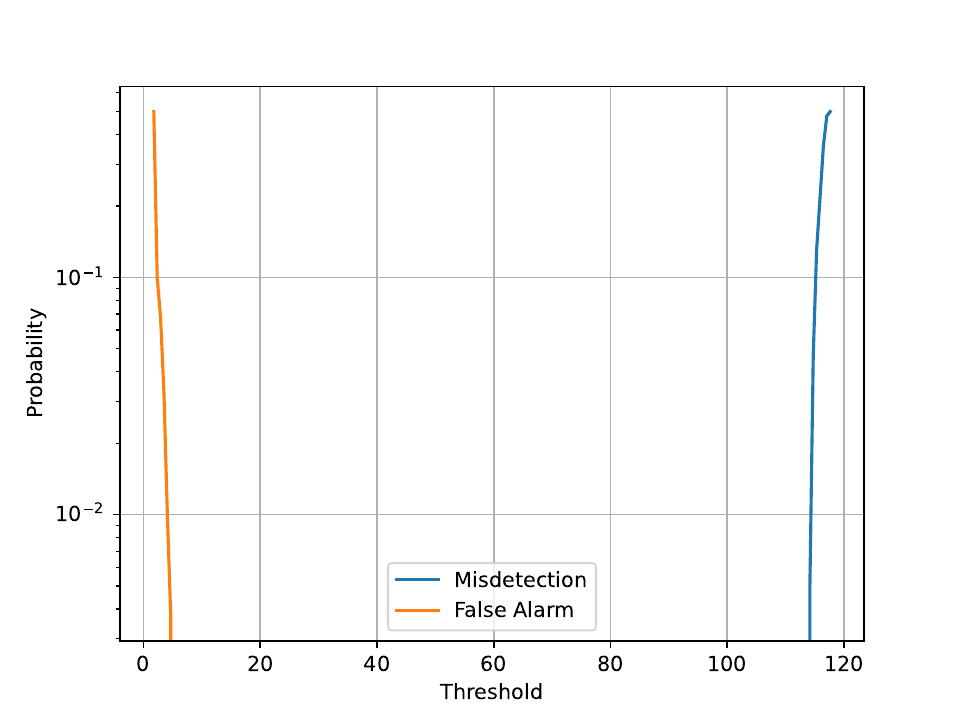}
    \caption{FA and MD probabilities as a function of the threshold $\tau$ for the Gaussian noise generator with the unsupervised learning approach.}
    \label{fig:unsup_error_gaussian}
\end{figure}

From Fig. \ref{fig:unsup_error_uniform} and \ref{fig:unsup_error_gaussian}, we can see how the model distinguishes perfectly between the jammed and not-jammed cases. This perfect detection is possible because the reconstruction error of the jammed case is approximately $50$ times higher than the reconstruction error of the case without jamming. Comparing Fig. \ref{fig:unsup_error_uniform} and \ref{fig:unsup_error_gaussian} shows no significant effect for the noise distribution of the jammer.

\begin{figure}
    \centering
    \includegraphics[width=1.0\hsize]{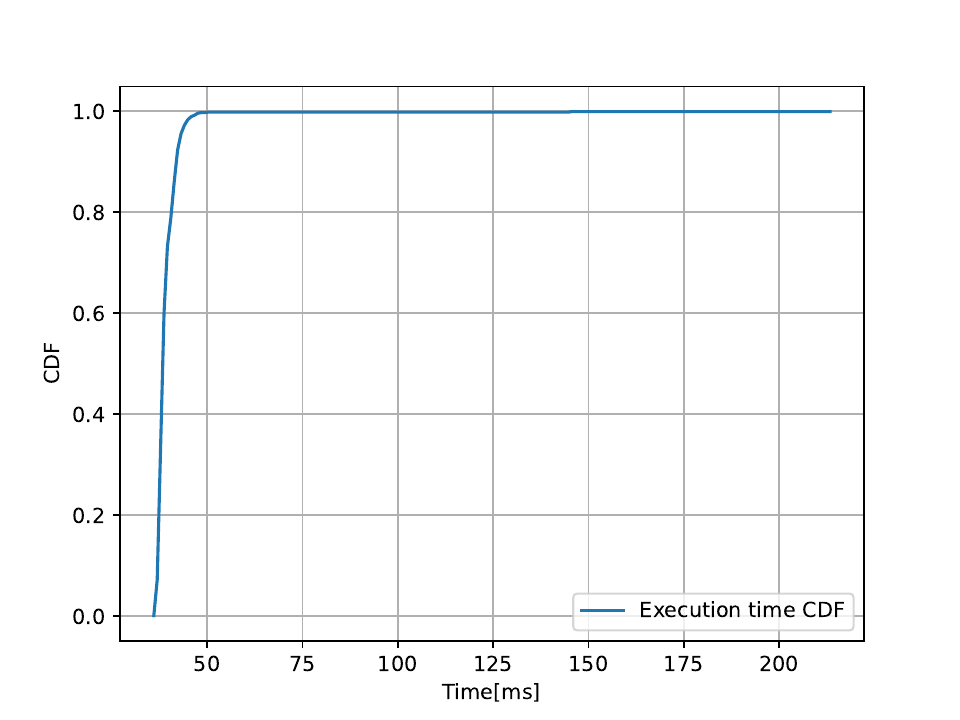}
    \caption{CDF of the classification time performed by the CAE with unsupervised learning: in $95\%$ of the cases it was below $48$ ms.}
    \label{fig:unsup_time}
\end{figure}

Fig. \ref{fig:unsup_time} plots the cumulative distribution function (CDF) of the classification time per sample for $1000$ trials. Each measurement was obtained using the CPU and includes the following steps:
\begin{itemize*}
    \item Loading into the memory the slice of data required to create a sample as described before.
    \item Creating the matrix through the PSD array computations and applying the monotonic function.
    \item Computing the output of the pre-trained model having as input the same sample.
    \item Comparing the output of the model to a given threshold using an \textbf{if} statement.
\end{itemize*}
This CDF allows to conclude that the classification time stays under $48$ ms in $95\%$ of the cases, allowing relatively fast detection and reaction to jamming attacks.

\subsection{Supervised Learning}
The training set was composed of $4500$ samples, equally distributed between the three cases: jammed, not-jammed and empty channel, not-jammed and ongoing transmission. Using the same distribution, the validation set was composed of $1800$ samples. This set is used to monitor the validation loss and stop the training as for the unsupervised learning process. The test set was composed of $1200$ samples, distributed in the same way as the training and validation set.

\begin{figure}
    \centering
    \includegraphics[width=1.0\hsize]{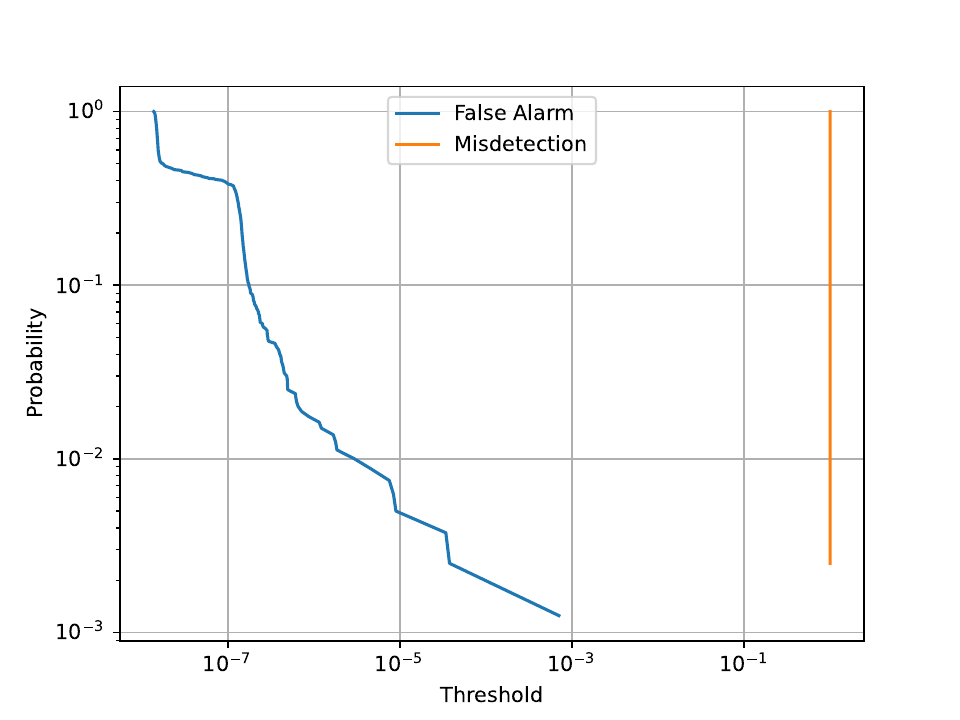}
    \caption{FA and MD probabilities as a function of the threshold $\tau\in [0,1]$ (with $y$ axis values normalized to $1$) for the uniform noise generator with the supervised learning approach.}
    \label{fig:sup_error_uniform}
\end{figure}

\begin{figure}
    \centering
    \includegraphics[width=1.0\hsize]{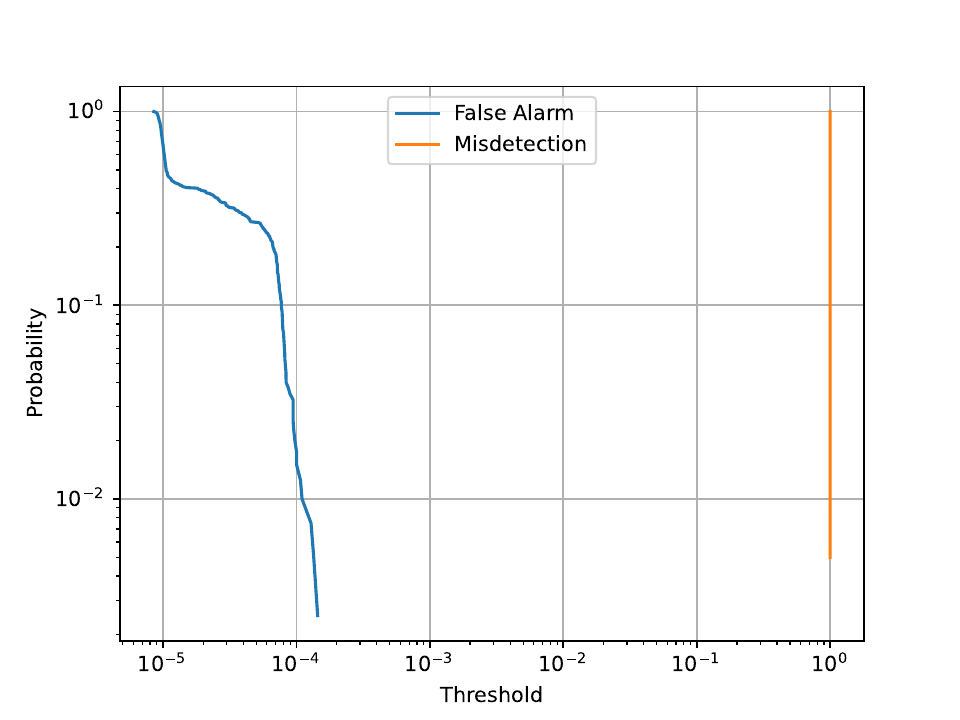}
    \caption{FA and MD probabilities as a function of the threshold $\tau\in [0,1]$ (with $y$ axis values normalized to $1$) for the Gaussian noise generator with the supervised learning approach.}
    \label{fig:sup_error_gaussian}
\end{figure}

Comparing the detection rates in Fig. \ref{fig:sup_error_uniform} and \ref{fig:sup_error_gaussian} to the  results in Section \ref{ssec:unsup_res}, shows that supervised learning reaches the highest accuracy. This becomes apparent by the absence of misdetection events and by the large threshold interval without false classification. This benefit of supervised learning, however, comes at a significant drawback that training is based on the signals of specific jamming attacks. Even slightly changing these signals may allow an attacker to evade the detection. Albeit showing slightly worse performance, the unsupervised learning model is not based on specific attacks but models not-jammed signals. A jamming attack is then detected as a significant deviation from this trusted state.

\begin{figure}
    \centering
    \includegraphics[width=1.0\hsize]{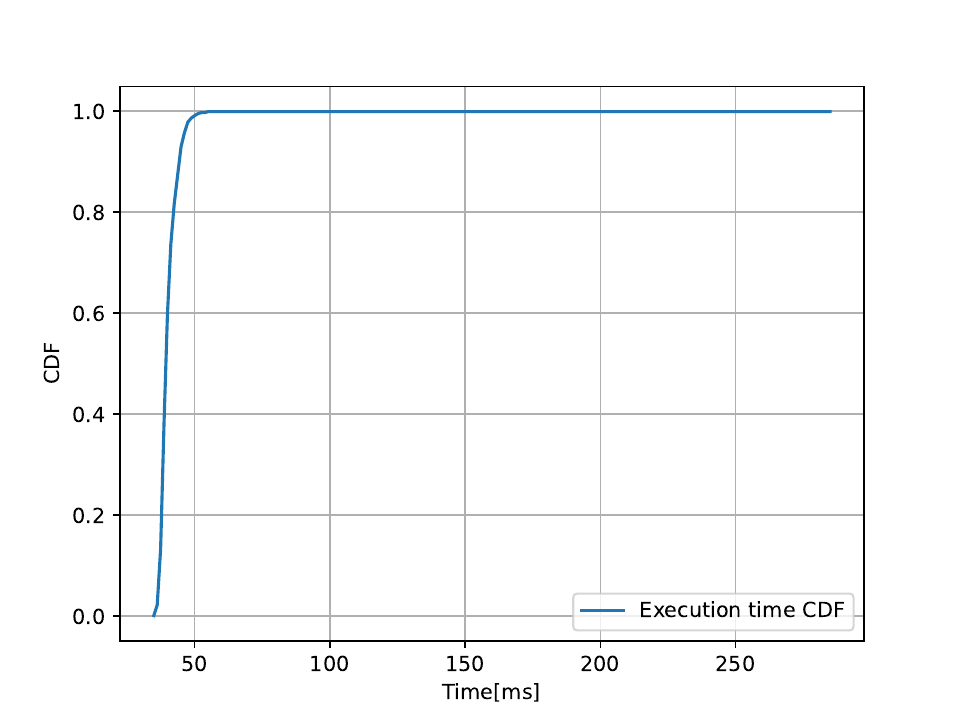}
    \caption{CDF of the classification time performed by the CNN with supervised learning: in 95\% of the
cases it was below $46$ ms.}
    \label{fig:sup_time}
\end{figure}

Fig. \ref{fig:sup_time} plots the CDF of the classification time per sample with the supervised learning model. The CDF is based on $1000$ trials. This result is similar to the classification time with unsupervised learning, despite the fact that the CNN (supervised) uses only half the parameters of the CAE (unsupervised). This suggests that a significant part of the computational complexity lies in the overhead of the ML model.

\section{Conclusions}
\label{sec:concl}
We proposed a method to detect jammers in 5G signals based on the PSDs of received radio signals. The method can be implemented as a separate network element (watchdog) and requires no interaction with the 5G system and no change of the 5G signals or standards. The computational complexity at the watchdog is low since measuring PSDs requires neither equalization nor synchronization. After stacking these PSDs to spectrograms, we constructed a trainable dataset avoiding the vanishing gradient problem. 

In our experiments, this approach shows very high accuracy at low computational complexity. CAE and CNN models both robustly distinguish between the jammed and not-jammed cases. The unsupervised CAE provides the additional benefit of being independent of the attacker. 

Based on these promising models, future work will cover a wider range of wireless scenarios and jammer operations. In particular, datasets for jamming the 5G Synchronization Signal Block (SSB) will be created and studied.
\balance
 
\bibliographystyle{IEEEtran}
\bibliography{conference}
\end{document}